# Large photocathode 20-inch PMT testing methods for the JUNO experiment

**N. Anfimov**[a] **on behalf of the JUNO collaboration.**

[a] *Joint Institute for Nuclear Research,
141980, 6 Joliot-Curie, Dubna, Russian Federation*
   *E-mail*: anphimov@gmail.com

ABSTRACT: The 20kt Liquid Scintillator (LS) JUNO detector is being constructed by the International Collaboration in China, with the primary goal of addressing the question of neutrino mass ordering (hierarchy). The main challenge for JUNO is to achieve a record energy resolution, ~3% at 1MeV of energy released in the LS, which is required to perform the neutrino mass hierarchy determination. About 20'000 large 20" PMTs with high Photon Detection Efficiency (PDE) and good photocathode uniformity will ensure an approximately 80% surface coverage of the JUNO detector. The JUNO collaboration is preparing equipment for the mass tests of all PMTs using 4 dedicated containers. Each container consists of 36 drawers. Each drawer will test a single PMT. This approach allows us to test 144 PMTs in parallel. The primary measurement in the container will be the PMT response to illumination of its photocathode by a low-intensity uniform light. Each of the 20000 PMTs will undergo the container test. Additionally, a dedicated scanning system was constructed for sampled tests of PMTs that allows us to study the variation of the PDE over the entire PMT photocathode surface. A sophisticated laboratory for PMT testing was recently built. It includes a dark room where the scanning station is housed. The core of the scanning station is a rotating frame with 7 LED sources of calibrated short light flashes that are placed along the photocathode surface covering zenith angles from the top of a PMT to its equator. It allows for the testing of individual PMTs in all relevant aspects by scanning the photocathode and identifying any potential problems. The collection efficiency of a large PMT is known to be very sensitive to the Earth Magnetic Field (EMF), therefore, understanding the necessary level of EMF suppression is crucial for the JUNO Experiment. A dark room with Helmholtz coils compensating the EMF components is available for these tests at a JUNO facility. The Hamamatsu R12860 20'' PMT is a candidate for the JUNO experiment. In this article the container design and mass-testing method, the scanning setup and scanning method are briefly described and preliminary results for performance test of this PMT are reported.

# Contents

## 1. The JUNO detector

The JUNO experiment is named after the Jiangmen Underground Neutrino Observatory [1]. JUNO is a multipurpose neutrino experiment designed to determine the neutrino mass hierarchy and precisely measure oscillation parameters by detecting reactor neutrinos, observing supernova neutrinos, studying atmospheric, solar, and geo-neutrinos, and by performing exotic searches. The JUNO detector is located in Kaiping, Jiangmen, in Southern China. It is about 53 km from the Yangjiang and Taishan nuclear power plants, both of which are under construction. The planned total thermal power of these reactors is 36 GW.

JUNO will use a spherical central detector with diameter of 35 m, which will be filled with 20 kton of LAB-based liquid scintillator and placed at 700 m underground. It will be equipped with ~ 20 000 20″ PMTs and with ~25 000 3″ PMTs. In order to perform a measurement of the neutrino mass hierarchy the JUNO detector must be able to discriminate fine oscillations in the observed inverse beta-decay spectrum. To do this, it is necessary to achieve a remarkable energy resolution of $\sigma$=3% at 1 MeV, which requires scintillation light collection of at least 1200 p.e./MeV. In order to obtain this level of light collection, the PMTs are required to have high and homogenous Photon Detection Efficiency (PDE) at a level of 30%.

### 1.1. JUNO 20'' Photomultiplier tubes.

The JUNO central detector readout will be made of up to 5k Hamamatsu (Japan) 20'' conventional dynode-PMTs and 15k NNVT (China) 20'' MCP-PMTs.

Table 1. Some of the main 20''-PMT's parameters requested for the JUNO central detector.

| Parameter | Min | Max | Typical | Unit |
|---|---|---|---|---|
| Photon Detection Efficiency at 420 nm | 24 | - | 27 | % |
| Gain | - | - | $10^7$ | - |
| PDE uniformity within photocathode area | - | 15 | - | % |
| Single photoelectron peak-to-valley ratio | 2.5 | - | 3.0 | - |
| Dark rate in operational mode | - | 50 | 20 | kHz |
| Insensitivity to magnetic field. | 0 | 5 | - | µT |

The Hamamatsu R12860 20'' PMT is one of the current candidates for the JUNO experiment. The JUNO collaboration has fixed very strict requirements for all PMTs. Some of the parameters are listed in table 1.

The JUNO detector will have Earth Magnetic Field (EMF) compensation to the level of <10% of the primary field, which at the JUNO location is about 50 µT. The last row in the table shows that each PMT should preserve its parameters within approved requirements in the residual magnetic field up to 5 µT.

## 2. Mass-testing methods

### 2.1 Container approach

All of the 20000 large-photocathode 20-inch PMTs will be tested in 4 containers which are designed and produced by the team from the University of Hamburg and the University of Tübingen. Each container is a 20' refrigerated container that can control the temperature within a range between -20°C and 45 °C with a precision of less than 1°C. In addition, the containers are lined with a multi-layer magnetic shielding based on silicon iron that guarantees a magnetic field of less than 10 % of the EMF in each of the 36 measurement positions.

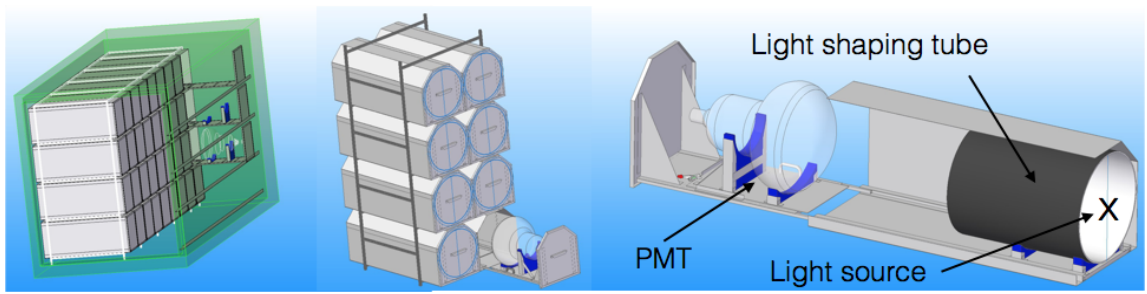

Fig.1. The Container: general view (left), drawers (middle), PMT layout in a drawer (right).

To provide identical and reproducible measurement conditions each container is equipped with 36 precision-made drawer boxes housed in a shelf, as can be seen in Figure 1. The complete shelf system (including the boxes) is made from aluminum, and nearly all surfaces inside the container and the drawer boxes are black. The PMTs are placed on removable trays equipped with a precise holder made from anti-static foam. The trays are then fixed to the drawer by a clamping lever, allowing for the precise positioning of the PMTs inside each drawer.

On the other side of the drawer boxes the light sources are mounted 50 cm away from the top of the PMT. The light sources are stabilized LEDs produced by the HVSYS company [2] , and are also used in the scanning stations. They are deployed behind optics (including a diffuser) designed to generate a suitable light field to illuminate the entire surface of the 20" PMTs with an intensity between 0.1 and 1.5 photons per LED pulse. A large light-shaping tube, coated black on one side and equipped with highly reflecting Tyvek on the other side, ensures that the sides of the PMTs are also illuminated (see Fig.1 right) . In addition, two of the four containers will be equipped with a picosecond Laser (wavelength of about 420 nm). The light of the lasers will be distributed by optical fibers producing a light field of similar intensity but within a narrower cone, optimized for TTS measurements.

Two of the four containers will be equipped with standard electronics in a VME crate outside the container. This allows for the testing of bare (naked) PMTs. Another test with the custom-made JUNO electronics is planned for the other two containers. Not all of the PMT characteristics can be tested inside the container. The container will mainly test the PDE and TTS of the PMTs, as well as the dark count rate and the pre- and after-pulse rates. It will also provide the optimal high voltage value for a gain of $10^7$ and the peak-to-valley (P/V) ratio for single p.e. measurements. Also, the full pulse shape information will be recorded, allowing for the determination of the average rise and fall times of the PMT pulses.

**2.2. Scanning station.**

The container approach is not sensitive to inhomogeneities of characteristics along the PMT's photocathode surface. Since all of the measurements in the container are performed in a constantly compensated magnetic field at the level of a few µT, it does not allow for the testing of a PMT's magnetic field sensitivity. In order to obtain these measurements a sampling of about a thousand PMTs will be tested more precisely. A special setup, called the scanning station, was designed and produced at the Joint Institute for Nuclear Research (see fig. 2). The scanning station is placed in a light-tight black room. In order to adjust/compensate for the EMF inside the black room Helmholtz coils are installed within the walls, floor and ceiling.

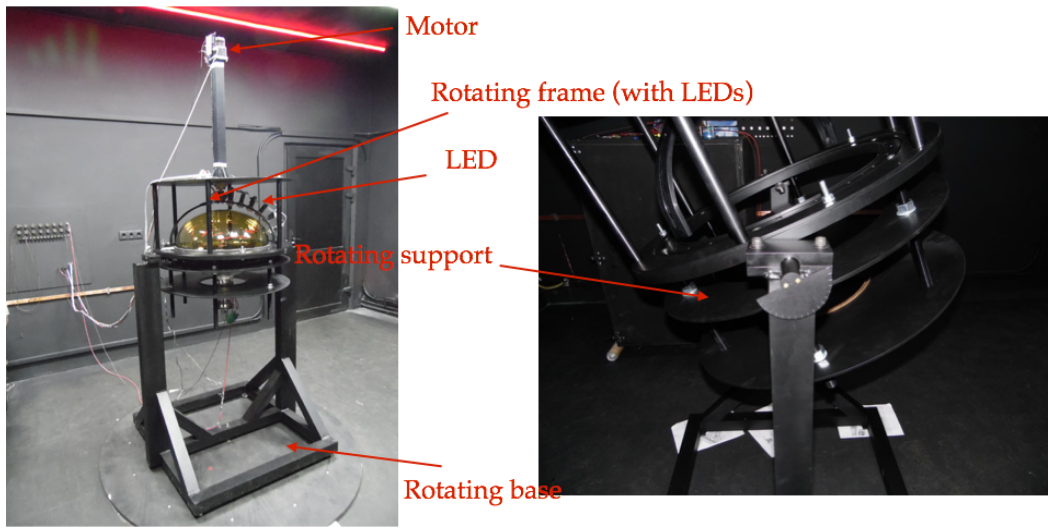

Fig.2. Scanning station general view in the dark room (left) and rotating support (right).

The core of the scanning station is a rotating frame with 7 stabilized compact pulsed light generators that are placed at different zenith angles. The frame is rotated by a step motor and covers all 360-deg azimuthal angles. A support system that holds the PMT allows for rotations in different spatial positions in order to put the PMT into different orientations with respect to the magnetic field provided by the dark room. It allows for the testing of individual

PMTs in all relevant aspects by scanning the photocathode, and allows for an in-depth understanding of the performance of a PMT, and may identify any potential problems.

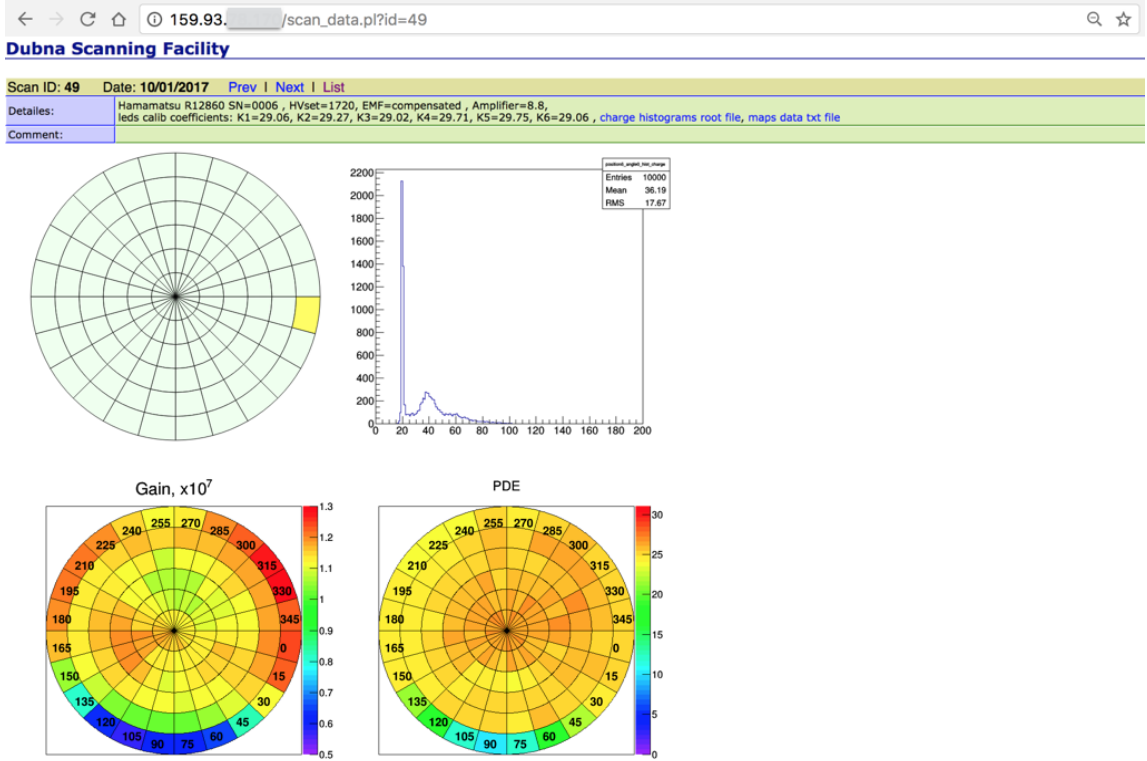

Fig.3. Scanning results for Hamamatsu R12860 HQE EA0006 presented by online database software

The testing method is based on very low-intensity light flashes [3] (~ 1 ph.e) to obtain gain, average number of photoelectrons, and other parameters. We are using a DRS4 [4] ADC to acquire PMT signals as a response to light flashes in the external trigger. Currently, software is set up to acquire 10k PMT signals, and then obtain a charge spectrum by integrating each of the waveforms. An example of a PMT charge spectrum is shown in fig.3 (top right). By using a calibrated light source we can characterize the photon detection efficiency (fig.3 bottom right) of the tested PMT.

Light sources are a speciality of the HVSYS company[2]. The light source used in this device is a stabilized, pulsed LED source, which is a compact device of 80x22x11 mm in size, implemented within a single package. The stabilization is provided by a PIN-photodiode that monitors the LED light. The PIN feedback is made up of an ADC digitizing PIN signal and a DAC controlling the LED amplitude. Both devices are embedded within a small chip of the microcontroller. A calibration of the light source is done using the calibrated photosensor, which is a small head-on-type linear-focused $1\frac{1}{8}$'' R1355 PMT with known QE (24% at 425 nm). Calibration of the PMT is done using a certified PIN photodiode. Because the LED source, which is under calibration, produces a 5 mm light spot in the middle of the calibration PMT photocathode, we assumed the collection efficiency of photoelectrons is close to 1, and, therefore, PDE ≅ QE. We can then define the average number of photons per LED flash.

The scanning procedure and data analysis are fully automatic. By analyzing histograms with a method similar to that described in [2] we obtain all of the main PMT parameters: gain, number of photoelectrons (PDE), single photoelectron resolution (P/V) etc. To present and store data a database with an online web interface was developed (fig.3). The user may choose any scanned PMT in the database and access each scanning point (fig.3 top left, yellow sector) on the PMT's surface, and check for spectrum consistency with typical charge spectra (fig.3. top right). If necessary, the user is able to download the entire charge histogram in root format and output all numerical values in text-file format. Data are represented in round-shape 2D-colored plots, which are called maps. Each circle on the map corresponds to one LED (zenith position) and each sector refers to an LED's azimuthal position. Numbers in edge sectors present azimuthal angle position.

## 3. Results of scanning for Hamamatsu R12860 PMTs

Two Hamamatsu R12860-50 PMTs with serial numbers EA0416, EA0418 were tested. We matched scanning results for both PMTs with Hamamatsu datasheets, which accompanied each PMT.

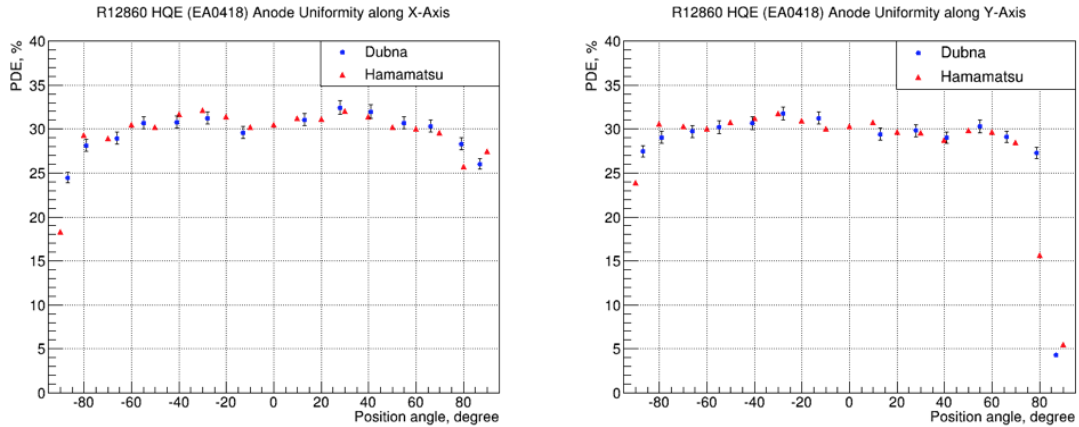

Fig.4. PDE vs PMT's zenith angle along X,Y axes for Hamamatsu R12860-50 EA0418.

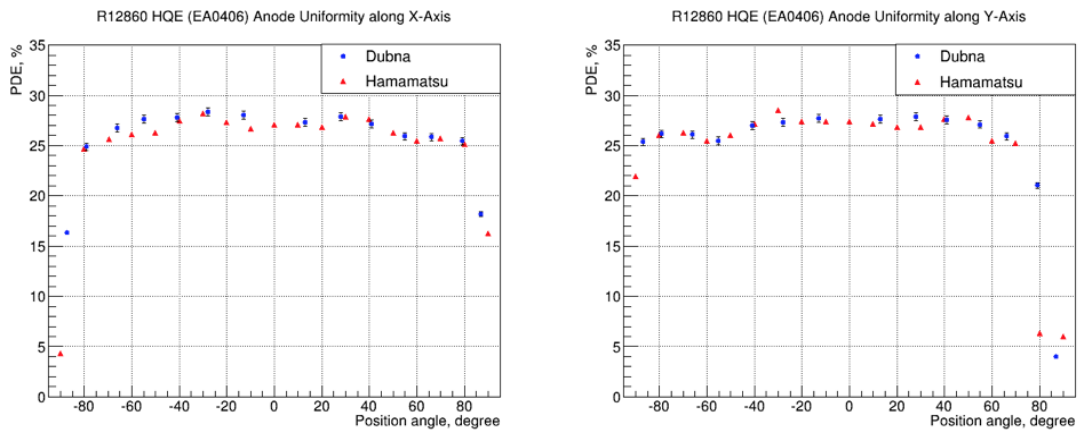

Fig.5. PDE vs PMT's zenith angle along X,Y axes for Hamamatsu R12860-50 EA0406.

The provided Hamamatsu Quantum Efficiency (QE) is measured from the top of the PMTs with a light spot of 30 mm diameter and Anode Uniformity along the X and Y axes, which is correlated with relative PDE. To perform the matching we assumed that the collection efficiency of photoelectrons for a large PMT is close to 1 for this spot, giving the approximation PDE ≅ QE. The anode uniformity was normalized in the center for the X and Y axis to the PDE (Hamamatsu on fig.4,5). In our measurements we oriented the PMT in such a way that the Hamamatsu X-axis corresponds to 0-180 azimuthal angle on the scanning maps in order to extract the PDE dependence for the X and Y axes (Dubna on fig.4,5) for the comparison. The results for both PMTs are shown in fig.4, 5. One can see that results match very well. Because large and small (calibration) PMTs have different geometries and different dynode systems, we may conclude that the assumption PDE ≅ QE for both cases is reasonable.

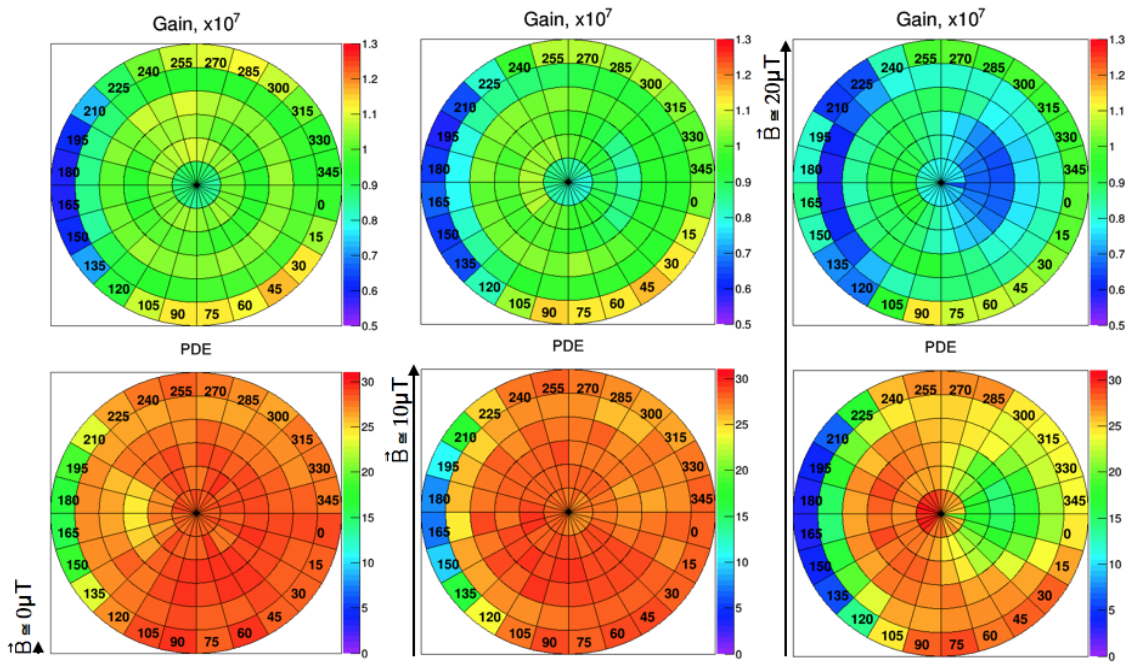

Fig.6. Hamamatsu R12860 HQE EA0006 PMT behavior in magnetic field: Compensated (left column), 10μT(middle column), 20 μT (right column). B vectors show direction and value of applied magnetic field with respect to the PMT.

The scanning station is an excellent instrument for studying the behavior of a PMT within a magnetic field. By adjusting currents in the Helmholtz coils we may vary the magnetic field vector. We oriented the PMT along the magnetic field vector. The magnetic field changes electron trajectories and has an influence on the collection efficiency (PDE) and Gain. Results for Hamamatsu R12860 HQE EA0006 with different strengths of magnetic field are shown in fig.6. The PMT operates well with 10 μT and start to fail the test requirement of Homogeneity of PDE > 15% and average PDE<24% with 20 μT. This study was performed with 6 LEDs. Edge circles (the outermost LED) on the maps are out of the guaranteed photocathode area (460 mm) and are not considered in the test.

**Acknowledgments**

We greatly appreciate Dr. Bjoern Soenke Wonsak for his help in the description of the container approach, and the JUNO collaboration members for their review and important corrections in the article. We would like to mention Dr. Zinoviy Krumshteyn, who has passed away, for his great contribution to this work.